\documentclass{article}
\usepackage{spconf}
\usepackage{tikz,tikz-qtree}	
\usepackage{graphicx}
\usepackage{amsmath,amssymb,bbm}
\usepackage{algorithm,algpseudocode}
\usepackage{url,multirow}

\newcommand{\bld}[1]{{\bf #1}}
\newcommand{\bs}[1]{\boldsymbol{#1}}
\newcommand{\Expectation}{\mathbb{E}}
\newcommand{\Probability}{\mathbb{P}}

\newcommand{\boundellipse}[4]
{(#1) ellipse [x radius = #2, y radius = #3, rotate = #4]}

\newcommand{\openRectangle}[5]
{ 				(#1,#2) 
                -- (#1,#2-#4) 
                -- (#1+#3,#2-#4)node[below,midway] {#5}
                -- (#1+#3,#2)}
\newcommand{\openRectangleDown}[5]
{ 				(#1,#2) 
                -- (#1,#2+#4) 
                -- (#1+#3,#2+#4)node[above,midway] {#5}
                -- (#1+#3,#2)}
                
\newcommand{\vArray}[3]
{ \matrix(m)[matrix of nodes,nodes in empty cells,
  nodes={inner sep=1pt,draw={#1},line width=0.2pt,text width=10pt, text height=10pt,text depth=0.1pt},
  row sep=-0.5\pgflinewidth,
  column sep=-\pgflinewidth,
  ]at (#2,#3){
    $\mu_1$\\
    $\mu_2$\\
     \\
     \\
     \\
     \\
     \\
     \\
   $\mu_K$\\
  };				}

\newcommand{\mobile}[2]
{
	\begin{scope}[shift={#1},rotate=#2,scale=0.7]		
		\draw[rounded corners=1pt,thick] (0,0) rectangle ++(0.3,0.5);
		\draw[] (0,0.1) -- (0.3,0.1);\draw[] (0,0.4) -- (0.3,0.4);
		\draw[] (0.125,0.05) -- (0.175,0.05);\draw[] (0.1,0.45) -- (0.2,0.45);
	\end{scope}
}
\title{Latent variable approach to diarization of audio recordings using ad-hoc randomly placed mobile devices}
\name{Srikanth~Raj~Chetupalli,
		Anirban~Bhowmick,
        and Thippur~V.~Sreenivas}
\address{Dept. of ECE, Indian Institute of Science, Bangalore, 560012.}

\begin{document}

\maketitle

\begin{abstract}
 Diarization of audio recordings from ad-hoc mobile devices using spatial information is considered in this paper. A two-channel synchronous recording is assumed for each mobile device, which is used to compute directional statistics separately at each device in a frame-wise manner. The recordings across the mobile devices are asynchronous, but a coarse synchronization is performed by aligning the signals using acoustic events, or real-time clock. Direction statistics computed for all the devices, are then modeled jointly using a Dirichlet mixture model, and the posterior probability over the mixture components is used to derive the diarization information. Experiments on real life recordings using mobile phones show a diarization error rate of less than $14\%$.
\end{abstract}
\begin{keywords}
Diarization, Dirichlet distribution, steered response power, acoustic sensor network, mobile devices.
\end{keywords}

\section{Introduction}
\label{sec:intro}
Consider a meeting scenario with several participants carrying mobile devices with one or more microphones. Mobile devices placed on a table can be considered to form an ad-hoc network of microphones. Spatial diversity provided by the multi microphone signals can be used to improve the performance of speech applications such as enhancement, recognition, diarization etc. However, such a setup is characterized by asynchronous recording at different devices, although microphones on the same device can record synchronously. A similar scenario is encountered in wireless acoustic sensor networks, where each node in the network can record using multiple microphones in a synchronous manner, but the signals at different nodes are asynchronous. In this paper, we first address the diarization task, i.e., ``who spoke when?" in an audio recording comprising of multiple sources (speakers), and signals recorded from an ad-hoc microphone array network.
\par Methods based on spectral features, spatial features or a combination of both are proposed for multi-channel diarization of audio recordings \cite{overview2006tranter, speaker2012anguera, review2012moattar}. In this paper, we consider the diarization of audio recordings using spatial features alone. Several solutions have been proposed utilizing spatial features, which use the 
time-difference-of-arrival (TDOA) features \cite{anguera2007acoustic, evans2009speaker, parada2017robust, nakamura2017blind}. However, the estimation of TDOA is sensitive to reverberation and noise. An alternate formulation based on a probabilistic spatial dictionary and Watson mixture modeling of directional features is proposed in \cite{ito2017data}. A pre-trained (data-driven) or pre-computed (physics based) spatial dictionary is used, which limits the application of the method to a finite set of source positions and known microphone geometry. Synchronous recording is assumed for all the microphones in the network in the above approaches. For the ad-hoc microphone network considered in this paper, microphones across the different mobile devices are asynchronous, and hence network-wide computation of TDOA or the directional features is not possible. However, it is possible to compute the directional features independently at each device, which can be combined later using a stochastic formulation.
\par Spatial response function computed using steered response power with the phase transform (SRP-PHAT) filtering is used as the spatialization measure. Assuming known microphone geometry at each device, the SRP response function is computed for a set of directions and these measures can be normalized to form a stochastic measure. We can use this as a ``directional statistic" feature and directional statistics of several devices can be combined using a latent variable mixture model. We  use a Dirichlet mixture model \cite{bishop2006pattern} for this purpose. The signals from different devices can be coarsely aligned using specific acoustic event such as a clap or a tap, or network time. Expectation-maximization \cite{dempster1977maximum} is then used for maximum likelihood estimation of the latent variables and the diarization information is derived from the posterior mixture component probabilities. Experiments on real life meetings recorded using commercial off-the-shelf mobile phones show that diarization error rate (DER) of less than $14\%$ is possible, even with coarse synchronization across the devices. 
\vspace{-5pt}
\section{Problem Formulation}
\begin{figure}[h]
	\centering
	\begin{tikzpicture}
		\draw[] (0,0.2) rectangle ++(6.5,3.7);
		\draw[rounded corners=5pt] (2,1.25) rectangle ++(3.5,1.5);

		\draw (1.5,2) ellipse (0.15 and 0.5);
		\draw (1.5,2) ellipse (0.15 and 0.2);
		\node[rotate=270] at (1.2,1.9) {\tiny speaker $\#1$};	
						
		\draw (3,3.2) ellipse (0.5 and 0.2);
		\draw (3,3.2) ellipse (0.15 and 0.2);
		\node[rotate=180] at (3.,3.55) {\tiny speaker $\#2$};	

		\draw (4.5,3.2) ellipse (0.5 and 0.2);
		\draw (4.5,3.2) ellipse (0.15 and 0.2);
		\node[rotate=180] at (4.5,3.55) {\tiny speaker $\#3$};	
		
		\draw (3,0.8) ellipse (0.5 and 0.2);
		\draw (3,0.8) ellipse (0.15 and 0.2);	
		\node[rotate=0] at (3.,0.45) {\tiny speaker $\#S$};	

		\draw (4.5,0.8) ellipse (0.5 and 0.2);
		\draw (4.5,0.8) ellipse (0.15 and 0.2);		
		
		\mobile{(2.7,1.4)}{70};
		\mobile{(4.7,1.5)}{40};
		\mobile{(3.0,2.2)}{20};
		\mobile{(4.5,2.4)}{120};
		\draw (2.85,1.4) circle [radius=0.1] node {\tiny $1$};	
		\draw (2.85,2.2) circle [radius=0.1] node {\tiny $2$};
		\draw (5.0,1.55) circle [radius=0.1] node {\tiny $P$};
	\end{tikzpicture}
	\vspace{-5pt}
	\caption{Meeting scenario}\label{fig:meetingScenario}
\end{figure}
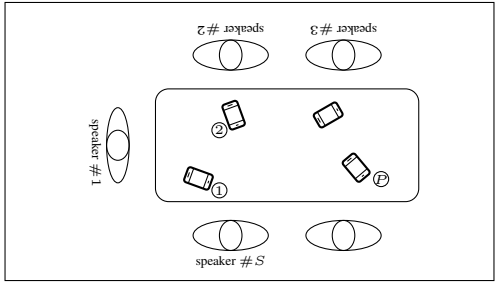

\vspace{-5pt}
\label{sec:ProblemFormulation}
Consider a meeting scenario with $S$ number of sources and $P$ number of mobile devices placed on a table as shown in \ref{fig:meetingScenario}. Let each device record a $M$-channel audio signal. Let $x_{m,p}[t]$ denote the speech signal recorded at the $m^{th}$ microphone of the $p^{th}$ device. Given the recordings at all the devices $\{x_{m,p}[t],\forall m\in[1~M];\forall p\in[1~P]\}$, the goal in this paper is to perform diarization, i.e., to identify ``who spoke when?" in the long conversation.
\par The mobile devices record the audio signal using their own independent clocks and no other external synchronization is used. However, the audio recordings across the different devices can be synchronized coarsely using the real-time log of network time or using acoustic events such as a tap or a clap. The synchronous recordings at a particular device provide us fpr beam steering computation of the source direction information (source direction statistics) independently at each mobile device. We consider the computation of directional statistics in a frame by frame manner. Each mobile device (randomly placed) forms its own SRP function and hence some what different directional statistics than other devices. We explore joint modeling of the directional statistics obtained at each of the devices using a latent variable mixture model. Even though the mobile devices are placed arbitrarily and the information about their own position and orientation is unknown (hence we cannot combine the individual directional statistics in a geometric formulation), we explore the power of stochastic modeling to derive the directional information. We note that in the proposed approach to diarization, the goal is not the exact position of the source, but to use the directional information to identify the source activity along the recorded time-line. We show that this is possible using a stochastic formulation of several mobile phone derived directional data.
\section{STATISTICAL DETECTION}
\label{sec:proposedMethod}
\subsection{Directional statistics features}
Let us consider steered response power (SRP) approach to compute the spatial features at each time-frame $n$ for each randomly placed mobile device separately. Let $\bld{x}[n,k]=\left[x_{1}[n,k] \dots x_{M}[n,k] \right]^T$ denote the multi-channel speech signal in the short time Fourier transform (STFT) domain for a microphone array. Let $\bld{a}[\theta,k]$ denote the steering vector corresponding to a source at a spatial direction $\theta$ for the frequency bin $k$ with respect to a local coordinate system centered at the array. Assuming free field propagation and a compact array, we have 
\begin{equation}
	\bld{a}[\theta,k]=\left[ 1~e^{\left(-\frac{j2\pi k \tau_{21}(\theta)}{K}\right)}\dots  e^{\left(-\frac{j2\pi k \tau_{M1}(\theta)}{K}\right)} \right]^T,
\end{equation}
where $\{\tau_{21}(\theta),\dots,\tau_{M1}(\theta) \}$ denote the TDOA values at the $M-1$ microphones with respect to the first microphone. SRP method \cite{dibiase2000high} can compute the spatial response function as,
\begin{equation}
	s[n,\theta]=\sum\limits_{k=1}^{K} \left|\bld{a}[\theta,k]^H \bld{x}_f[n,k] \right|^2,
\end{equation}
where $\bld{x}_f[n,k]=\frac{\bld{x}[n,k]}{|\bld{x}[n,k]|}$ is the signal phase vector obtained after PHAT filtering.
\par The response function $s[n,\theta]$ is evaluated at $L$ discrete angular positions $\bs{\Theta}=\{\theta_1,\dots,\theta_L\}$ with respect to the array. Since the source can be assumed to be relatively stationary compared to STFT/SRP computation, we smooth the discrete SRP function across time using recursive averaging,
\begin{equation}
	\tilde{s}[n,\theta_l]=\alpha \tilde{s}[n,\theta_l] + (1-\alpha) s[n-1,\theta_l].
\end{equation}
Smoothed SRP function is then normalized to represent the estimated directional statistics which is then used as feature for the mixture density modeling. 
\par Let $\bld{s}[n] \triangleq \frac{1}{C} \left[\tilde{s}[n,\theta_1]\dots \tilde{s}[n,\theta_L] \right]^T$, where $C=\sum\limits_{l=1}^{L} \tilde{s}[n,\theta_l]$ is the normalization constant. Thus the vector $\bld{s}[n]$ is a positive function and sums up to unity; hence, we can interpret the $\bld{s}[n]$ as a PMF over the set $\bs{\Theta}$ at each time-frame $n$. 
\par In the present formulation, we compute the directional statistics independently for each mobile device, and obtain $P$ number of directional statistic features $\{\bld{s}_p[n]\}$, one per mobile device, at each time frame. 
However, due to reverberation in the enclosure and other recording noise, $\{\bld{s}_p[n]\}$ do have estimation errors and hence a further statistical formulation is required to effectively combine the information from several recording devices. 
\subsection{Latent variable joint modeling}
\begin{figure}[h]
	\centering
	\begin{minipage}[b]{0.48\linewidth}
		\centering
		\begin{tikzpicture}[scale=0.8,every node/.style={scale=0.8}]
			\draw[rounded corners=5pt] (3.8,-0.5) rectangle ++(2.5,4.0);
			\draw[rounded corners=5pt] (4.0,0) rectangle ++(2.0,2.0);
			\draw (5,4.1) node (priorParameters){$\{\pi_s\}$};
			\node[draw,circle] at (5,2.8) (selectionVector) {$\bld{z}_n$};		
			\node[draw,circle,fill=gray] at (5,0.8) (observations) {$x_{m,p}[t]$};		
			\draw[->] (priorParameters) -- (selectionVector);
			\draw[->] (selectionVector) -- (observations);
			\draw (6.0,-0.3) node (n){$N$};
			\draw (5.75,0.2) node (n){\scriptsize $MP$};
		\end{tikzpicture}
		\centerline{(a)}
	\end{minipage}
	\begin{minipage}[b]{0.48\linewidth}
		\centering
		\begin{tikzpicture}[scale=0.8,every node/.style={scale=0.8}]
			\draw[rounded corners=5pt] (3.8,-0.5) rectangle ++(2.5,4.0);
			\draw[rounded corners=5pt] (4.0,0) rectangle ++(2.0,2.0);
			\draw (5,4.1) node (priorParameters){$\{\pi_s\}$};
			\node[draw,circle] at (5,2.8) (selectionVector) {$\bld{z}_n$};		
			\node[draw,circle,fill=gray] at (5,0.8) (observations) {$\bld{s}_p[n]$};		
			\draw[->] (priorParameters) -- (selectionVector);
			\draw[->] (selectionVector) -- (observations);
			\draw (6.0,-0.3) node (n){$N$};
			\draw (5.75,0.2) node (n){\scriptsize $P$};
		\end{tikzpicture}	
		\centerline{(b)}		
	\end{minipage}
		\vspace{-5pt}
	\caption{Generative model of (a) the microphone observations and (b) the directional statistics.}\label{fig:graphicalModel}
\end{figure}
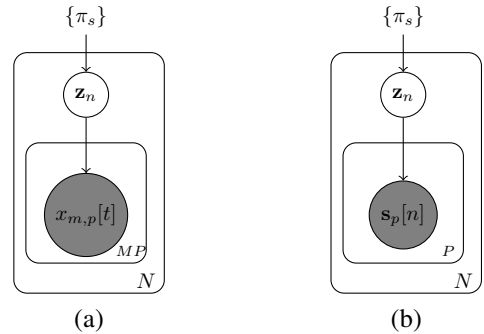
We model the set of distributions $\{\bld{s}_p[n]\}, 0\leq n \leq N-1$ jointly using a mixture model. A graphical model describing the generation of observations is shown in Fig. \ref{fig:graphicalModel}. The latent variable selection vector $\bld{z}_n$ selects a directional position (hence a source or a speaker), from a set of $S$ sources based on a Bernoulli distribution with parameters $\bs{\pi}$, i.e., $\Probability(\bld{z}_n|\bs{\pi})=\prod\limits_{s=1}^{S} \pi_s^{z_{ns}}$.
Now the overall generative model of the statistical observations can be stated as follows: the signal from a selected direction/speaker results in the observed signals $\{x_{m,p}(t)\}$ at the microphones of the devices, or equivalently the derived independent directional statistic features $\{\bld{s}_p[n]\}$ at the $P$ number of devices, according to
\begin{equation}
	\Probability(\{\bld{s}_p[n]\} |z_{ns}=1,\bs{\Delta})=\prod\limits_{p=1}^{P} \Probability(\bld{s}_p[n] |\bs{\delta}_{sp}),
\end{equation}
where $\bs{\Delta}=\{\bs{\delta}_{sp},\forall s,p\}$ is the set of parameters of all the distributions. A Dirichlet distribution \cite{bishop2006pattern} is used to model the directional statistics, to suit the discrete nature of the directional data and to suit the EM derivation. Hence,
\begin{equation}
	\Probability\left(\bld{s}_p[n] \big| \bs{\delta}_{sp}\right)=\mathcal{D}( \bld{s}_p[n];\bs{\delta}_{sp}),
\end{equation}
where standard Dirichlet distribution has the form,
\begin{equation}
	\mathcal{D}( \bld{s}_p[n];\bs{\delta}_{sp})= \frac{\Gamma\left(\sum\limits_{l=1}^{L} \delta_{sp}[l]\right)}{\prod\limits_{l=1}^{L} \Gamma \left( \delta_{sp}[l] \right) }\prod\limits_{l=1}^{L} \bld{s}_p[n,l]^{\delta_{sp}[l]-1}.
\end{equation}
We assume the directional data to be independent across time, which results in the model, 
\begin{equation}
	\Probability(\bld{S},\bld{Z}|\bs{\Delta},\bs{\pi})=\prod\limits_{n=0}^{N-1}\prod\limits_{s=1}^{S}\left[\pi_s \prod\limits_{p=1}^{P} \mathcal{D}(\bld{s}_p[n] |\bs{\delta}_{sp}) \right]^{z_{ns}},
\end{equation}
The parameters $\bs{\Delta}$ and $\bs{\pi}$ are estimated by maximizing the total likelihood function using the expectation-maximization (EM) algorithm. At iteration-$i$, the EM algorithm involves computation of (i) the posterior distribution $\Probability\left(\bld{Z}|\bld{S},\bs{\Delta}^{(i)},\bs{\pi}^{(i)} \right)$, and (ii) maximization of the expected joint likelihood objective $Q(\bs{\Delta},\bs{\pi})=\Expectation \{ \log \Probability(\bld{S},\bld{Z}|\bs{\Delta},\bs{\pi}) \}$.
\par It can be shown that, $\Probability\left(\bld{Z}|\bld{S},\bs{\Delta}^{(i)},\bs{\pi}^{(i)} \right)$ is an independent Bernoulli distribution with parameter,
\begin{equation}
	\Probability \left(z_{ns}=1|\{\bld{s}_p[n]\},\bs{\Delta}^{(i)},\bs{\pi}^{(i)} \right) =  \frac{\pi_s^{(i)} \prod\limits_{p=1}^{P} \mathcal{D}(\bld{s}_p[n];\bs{\delta}_{sp}^{(i)}) }{\sum\limits_{s=1}^{S}\pi_s^{(i)} \prod\limits_{p=1}^{P} \mathcal{D}(\bld{s}_p[n];\bs{\delta}_{sp}^{(i)}) }
\end{equation}
and $\Expectation\{z_{ns}\}\triangleq \gamma_{ns}^{(i+1)} = \Probability(z_{ns}=1 \big|\{\bld{s}_p[n]\},\bs{\Delta}^{(i)},\bs{\pi}^{(i)})$.
\par In the maximization step, the function $Q(\bs{\Delta},\bs{\pi})$ is maximized:
\begin{multline}\label{eqn:qFunction}
	Q(\bs{\Delta},\bs{\pi})=\sum\limits_{n=0}^{N-1} \sum\limits_{s=1}^{S} \gamma_{ns}^{(i)} \log \pi_s+\\
	\sum\limits_{n=0}^{N-1} \sum\limits_{s=1}^{S} \gamma_{ns}^{(i)} \sum\limits_{p=1}^{P} \log \mathcal{D}(\bld{s}_p[n];\bs{\delta}_{sp}).
\end{multline}
Maximization of eqn. \eqref{eqn:qFunction} with respect to $\pi_s$ subject to the constraint $\sum\limits_{s=1}^{S} \pi_s=1$ results in the estimate,
\begin{equation}
	{\pi}_s^{(i+1)}=\frac{N_s}{N},\mbox{~~where~~}N_s=\sum\limits_{n=0}^{N-1} \gamma_{ns}^{(i+1)}.	
\end{equation}
Maximization of \eqref{eqn:qFunction} with respect to $\bs{\delta}_{sp}$ requires solving the problem:
\begin{equation}
	{\bs{\delta}}_{sp}^{(i+1)}=\underset{\delta_{sp}}{\arg\max} \sum\limits_{n=0}^{N-1} \gamma_{ns}^{(i+1)} \log \mathcal{D}(\bld{s}_p[n] ;\bs{\delta}_{sp}).
\end{equation}
Substituting for $\mathcal{D}(\bld{s}_p[n] ;\bs{\delta}_{sp})$, we get the optimization problem as,
\begin{multline}
	{\bs{\delta}}_{sp}^{(i+1)}= \underset{\bs{\delta}_{sp}}{\arg\max}
	\sum\limits_{n=0}^{N-1} \gamma_{ns}^{(i+1)} \left[ \log \Gamma\left(\sum\limits_{l=1}^{L} \delta_{sp}[l]\right) - \right.\\\left. \sum\limits_{l=1}^{L} \log \Gamma(\delta_{sp}[l]) + \sum\limits_{l=1}^{L} \left(\delta_{sp}[l]-1 \right) \log \bld{s}_{p}[l]  \right].
\end{multline}
Gradient-descent based algorithm is used to solve for $\{{\bs{\delta}}_{sp},~\forall~s,p\}$ as shown in \cite{minka2000estimating}.
\subsection{Diarization}
At convergence of the EM algorithm, the posterior parameter, $\gamma_{ns}^{*}$ denotes the probability of $s^{th}$ source being active at $n^{th}$ time frame. The diarization information is obtained as the source label $s$ at each time frame $n$ using the max-rule over $s$,
\begin{equation}
	\hat{s}[n]=\underset{s}{\arg\max}~\gamma_{ns}^{*}.
\end{equation}
\section{Experiments and Results}
\label{sec:experimentsAndResults}
 Real-life meeting recordings are used for the evaluation of the proposed scheme. Three mobile phones are placed in an arbitrary orientation on a table in a reverberant enclosure (RT60 $\approx650$ ms). Each mobile is configured to record stereo signals at $F_s=48~KHz$. The recorded signals are down-sampled to $16$ KHz to confine STFT to $8~KHz$. The sound from a tap on the table is used as the acoustic event to align the signals across the mobile devices. We consider five recordings with three participants in each recording for evaluating diarization. The participants are chosen from three male speakers and one female speaker. The duration of the recordings varied from $5$ minutes to $10$ minutes, and the recordings are annotated at the speaker level manually. The mobile phones and participants are placed randomly for all the five recordings. 
\par STFT analysis is carried out using frames of size $64~ms$ with $50\%$ overlap between successive frames. In the SRP-PHAT computation, the beam steering is performed with a resolution of $4^o$ ($L=46$). Computation of the steering vector used in SRP-PHAT requires knowledge of the spacing between the microphones. For the commercial devices used in this experiment, since exact spacing is unknown, we choose a maximum spacing of $0.16~m$. This will affect only the local angle $\theta_l$ and does not alter the probability measures. The parameter $\alpha$ used for obtaining smooth directional statistics is chosen to be $0.9$. EM algorithm for DMM estimation is initialized using the method suggested in \cite{minka2000estimating}, and the maximum number of iterations is chosen to be $100$. The number of sources $S$ is assumed to be known in this experiment. However, it is possible to estimate the number of sources, by using the histogram of the peak locations of directional statistics features. The diarization performance is measured using DER, and computed using the NIST speech recognition scoring toolkit\footnote{\url{https://www.nist.gov/itl/iad/mig/tools}}, with a collar interval of $0.25~s$. The proposed algorithm assigns each frame to a single source, an estimate of the oracle performance is obtained using ground truth labels where we assigned the label of previous frame to frames with speaker overlap.
\par
\begin{figure}[!h]
	\centering
	\includegraphics[width=0.95\linewidth,height=5in,trim={0.45in 0 0 0},clip]{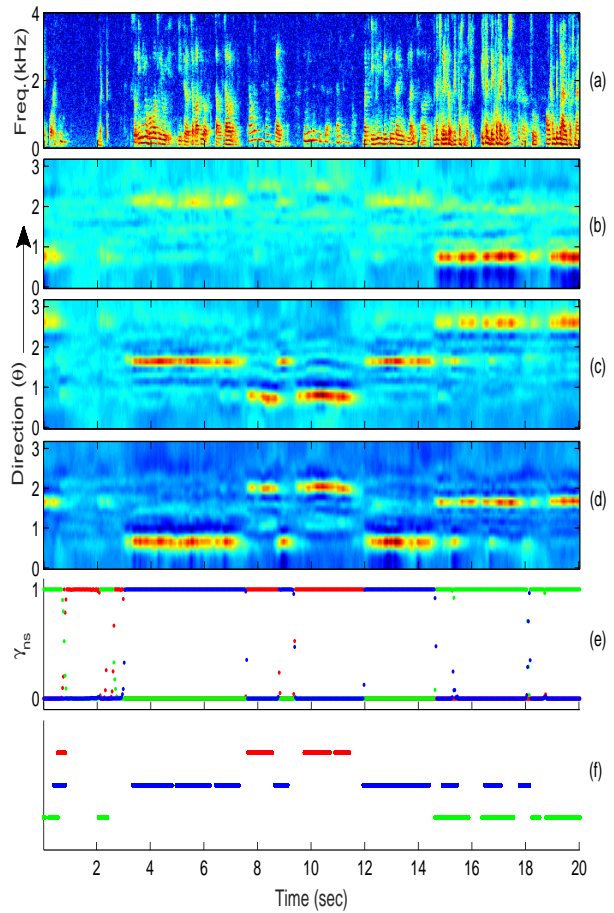}
	\vspace{-10pt}
	\caption{(a) Spectrogram of a microphone signal, and (b,c,d) computed directional statistics $\{\bld{s}_p[n]\}$ for the three mobile devices, (e) estimated posterior speaker probability $\{\gamma_{ns},~s=1,2,3\}$ shown in respective color, and (f) ground truth source activity denoted by three colors.}\label{fig:illustrationOfSpec}
	\vspace{-5pt}
\end{figure}

 Fig. \ref{fig:illustrationOfSpec} shows an illustration of the directional statistic features computed at the three mobile devices along with the spectrogram of the speech recorded at one of the devices for one of the recordings (illustrations for all the recordings are available here\footnote{\url{http://www.ece.iisc.ernet.in/~sraj/mDiar.html}}).
We see that the spatial features of the sources differ at the three devices, and the discriminability between two source positions is also different. However, there is one-one correspondence between the feature patterns across the devices. For example, in the first mobile recording, the directional statistics contain a clear peak only for one of the sources, and the energy is less directional for the other two sources. This may be due to the directionality and placement of the microphones in the mobile device. However, joint modeling along with the other devices helps in estimating the correct source regions. Source posterior $\{\gamma_{ns} \forall n\}$ is shown in Fig. \ref{fig:illustrationOfSpec}(e). We see that estimated speaker activity closely matches the ground truth shown in Fig. \ref{fig:illustrationOfSpec}(f). We note that, silence regions and also segments with overlapped speakers are assigned to the previous segmented speaker. This is because of the smoothing step in the computation of directional statistics. 
\begin{table}[h]
\centering
\vspace{-5pt}
\caption{DER performance (\%) for five recordings $R1-R5$}\label{tab:performance}
\begin{tabular}{|c|c|c|c|c|c|c|}
\hline
ID       & R1   & R2   & R3   & R4   & R5  & Avg.  \\ \hline
Proposed & 13.1 & 12.5 & 20.9 & 14.0 & 6.5 & 13.4 \\ \hline
Oracle   & 11.3 & 10.8 & 20.5 & 13.7 & 5.6 & 12.4 \\ \hline
\end{tabular}
\end{table}
\par Performance of the proposed algorithm on the five recorded conversations is shown in Tab. \ref{tab:performance}. The performance varies across the different recordings, due to the different source and microphone placements, and different amounts of overlap between the sources during the conversation. DER is found to be high for some conversations with more overlap. However, for all the recordings, the performance of the proposed algorithm is with in $2\%$ from the oracle performance.

\section{Conclusions}
\label{sec:conclusions}
 Coarse synchronization of different mobile devices and joint modeling of directional statistics computed per device is found to be sufficient for identifying "who spoke when?" in audio recordings from randomly placed mobile devices. This is true despite the unknown variabilities such as the nature of the microphones, their orientation within different mobile devices and also the random placement of the mobile devices. Presently a single source is assigned for each time-frame, but the method can be extended to predict multiple source activity, which can further improve the diarization performance.


\newpage
\bibliographystyle{IEEEtran}
\bibliography{references}

\begin{thebibliography}{10}
\providecommand{\url}[1]{#1}
\csname url@samestyle\endcsname
\providecommand{\newblock}{\relax}
\providecommand{\bibinfo}[2]{#2}
\providecommand{\BIBentrySTDinterwordspacing}{\spaceskip=0pt\relax}
\providecommand{\BIBentryALTinterwordstretchfactor}{4}
\providecommand{\BIBentryALTinterwordspacing}{\spaceskip=\fontdimen2\font plus
\BIBentryALTinterwordstretchfactor\fontdimen3\font minus
  \fontdimen4\font\relax}
\providecommand{\BIBforeignlanguage}[2]{{%
\expandafter\ifx\csname l@#1\endcsname\relax
\typeout{** WARNING: IEEEtran.bst: No hyphenation pattern has been}%
\typeout{** loaded for the language `#1'. Using the pattern for}%
\typeout{** the default language instead.}%
\else
\language=\csname l@#1\endcsname
\fi
#2}}
\providecommand{\BIBdecl}{\relax}
\BIBdecl

\bibitem{overview2006tranter}
S.~E. Tranter and D.~A. Reynolds, ``An overview of automatic speaker
  diarization systems,'' \emph{IEEE Trans. Audio, Speech, Lang. Process.},
  vol.~14, no.~5, pp. 1557--1565, Sept 2006.

\bibitem{speaker2012anguera}
X.~Anguera, S.~Bozonnet, N.~Evans, C.~Fredouille, G.~Friedland, and O.~Vinyals,
  ``Speaker diarization: A review of recent research,'' \emph{IEEE Trans.
  Audio, Speech, Lang. Process.}, vol.~20, no.~2, pp. 356--370, Feb 2012.

\bibitem{review2012moattar}
\BIBentryALTinterwordspacing
M.~Moattar and M.~Homayounpour, ``A review on speaker diarization systems and
  approaches,'' \emph{Speech Communication}, vol.~54, no.~10, pp. 1065 -- 1103,
  2012. [Online]. Available:
  \url{http://www.sciencedirect.com/science/article/pii/S0167639312000696}
\BIBentrySTDinterwordspacing

\bibitem{anguera2007acoustic}
X.~Anguera, C.~Wooters, and J.~Hernando, ``Acoustic beamforming for speaker
  diarization of meetings,'' \emph{IEEE Trans. Audio, Speech, Lang. Process.},
  vol.~15, no.~7, pp. 2011--2022, Sept 2007.

\bibitem{evans2009speaker}
N.~W.~D. Evans, C.~Fredouille, and J.~Bonastre, ``Speaker diarization using
  unsupervised discriminant analysis of inter-channel delay features,'' in
  \emph{Proc. Int. Conf. Acoust., Speech, Signal Process. (ICASSP)}, April
  2009, pp. 4061--4064.

\bibitem{parada2017robust}
P.~P. Parada, D.~Sharma, T.~van Waterschoot, and P.~A. Naylor, ``Robust
  statistical processing of tdoa estimates for distant speaker diarization,''
  in \emph{Proc. European Signal Process. Conf. (EUSIPCO)}, Aug 2017, pp.
  86--90.

\bibitem{nakamura2017blind}
K.~Nakamura and T.~Mizumoto, ``Blind spatial sound source clustering and
  activity detection using uncalibrated microphone array,'' in \emph{2017 25th
  European Signal Processing Conference (EUSIPCO)}, Aug 2017, pp. 2438--2442.

\bibitem{ito2017data}
N.~Ito, S.~Araki, and T.~Nakatani, ``Data-driven and physical model-based
  designs of probabilistic spatial dictionary for online meeting diarization
  and adaptive beamforming,'' in \emph{Proc. European Signal Process. Conf.
  (EUSIPCO)}, Aug 2017, pp. 1165--1169.

\bibitem{bishop2006pattern}
C.~M. Bishop, \emph{Pattern recognition and machine learning}.\hskip 1em plus
  0.5em minus 0.4em\relax springer, 2006.

\bibitem{dempster1977maximum}
A.~P. Dempster, N.~M. Laird, and D.~B. Rubin, ``Maximum likelihood from
  incomplete data via the em algorithm,'' \emph{Journal of the royal
  statistical society. Series B (methodological)}, pp. 1--38, 1977.

\bibitem{dibiase2000high}
J.~H. DiBiase, \emph{A high-accuracy, low-latency technique for talker
  localization in reverberant environments using microphone arrays}.\hskip 1em
  plus 0.5em minus 0.4em\relax Brown University, Providence RI, USA, May 2000.

\bibitem{minka2000estimating}
T.~Minka, ``Estimating a dirichlet distribution,''
  \url{https://tminka.github.io/papers/dirichlet/minka-dirichlet.pdf}, 2012,
  [Online; accessed 28-October-2018].

\end{thebibliography}

\end{document}